# Efficiently Searching for Frustrated Cycles in MAP Inference

**David Sontag, Do Kook Choe, Yitao Li**
Department of Computer Science
Courant Institute of Mathematical Sciences
New York University

## Abstract

Dual decomposition provides a tractable framework for designing algorithms for finding the most probable (MAP) configuration in graphical models. However, for many real-world inference problems, the typical decomposition has a large integrality gap, due to frustrated cycles. One way to tighten the relaxation is to introduce additional constraints that explicitly enforce cycle consistency. Earlier work showed that cluster-pursuit algorithms, which iteratively introduce cycle and other higher-order consistency constraints, allows one to exactly solve many hard inference problems. However, these algorithms explicitly enumerate a candidate set of clusters, limiting them to triplets or other short cycles. We solve the search problem for cycle constraints, giving a nearly linear time algorithm for finding the most frustrated cycle of arbitrary length. We show how to use this search algorithm together with the dual decomposition framework and cluster-pursuit. The new algorithm exactly solves MAP inference problems arising from relational classification and stereo vision.

## 1 Introduction

Graphical models such as Markov random fields have been successfully applied to many fields, from computer vision and natural language processing, to computational biology. Exact probabilistic inference is generally intractable in complex models having many dependencies between the variables. Here we consider the problem of finding the most probable (MAP) assignment in graphical models with discrete states.

Dual decomposition provides a powerful framework for designing tractable MAP inference algorithms [19]. This approach attempts to minimize an upper bound on the MAP assignment by reparameterization of the potentials. Many optimization algorithms have been proposed to minimize the dual, such as those based on subgradients [10, 13], message-passing approaches based on block coordinate-descent [6, 11, 14, 23], and provably convergent algorithms [7, 9, 16]. Every decomposition corresponds to a particular linear programming (LP) relaxation. The dual LP that is most frequently solved by these algorithms corresponds to the pairwise LP relaxation, which enforces local consistency constraints between factors that share variables.

However, for many real-world inference problems, the pairwise LP relaxation has a large integrality gap. These situations arise because there are frustrated cycles where a fractional solution can obtain a higher objective value by letting the edge marginals along the cycle be globally inconsistent. We can try to avoid these fractional solutions by instead optimizing over a tighter LP relaxation. There are well-known methods for tightening relaxations, such as the Sherali-Adams hierarchy of relaxations which uses cluster consistency constraints to enforce the consistency of all edge marginals in a cluster, for all clusters of some fixed size [15]. However, these linear programs are typically computationally infeasible to solve because they tighten the relaxation uniformly across all of the problem. For graphical models of moderate size, even the first lifting of the Sherali-Adams hierarchy (all triplet clusters, also called the *cycle relaxation*) is too slow to optimize over.

Several authors have proposed cluster-pursuit algorithms that iteratively tighten the dual using cycle consistency constraints [3, 9, 12, 21, 24]. These approaches are based on dictionary enumeration. They consider a small set of candidate cycles (e.g., all 3-cycles in a graph, or 4-cycles for grids), and evaluate a score to decide which cycle clusters to add to the decomposition. This explicit enumeration limits the applicability of these algorithms to graphical models where it is obvious where to look for the candidate cycles. However, many difficult inference problems are on sparse graphical models with few short cycles, such as the factor graphs used for low-density parity check codes, or Markov random fields whose structure follows that of a social network or the web graph (frequently used for relational classification).

In this paper, we address the most significant shortcoming of these cluster-pursuit algorithms: the difficulty of finding where to tighten the relaxation in sparse graphs. We show in Section 4 that, for non-binary graphical models, it is NP-hard to find the best cycle according to the bound criterion score used in earlier work [21]. Thus, we need an alternative approach to address the search problem of finding where to tighten the relaxation.

We describe a new approach where, as before, we solve the dual of the pairwise LP relaxation, but where we now search for $k$-ary cycle inequalities [20] to tighten the relaxation rather than cluster consistency constraints. We consider the same greedy bound minimization criterion used earlier for cluster consistency constraints, corresponding to the amount that the dual objective would decrease with just one coordinate descent step on the new dual variable(s). We show that one can, in near-linear time (in the size of the projection graph), find a $k$-ary cycle inequality whose guaranteed bound decrease is a constant fraction of the best possible bound decrease according to this criterion. The resulting algorithm is similar to the separation algorithm given in [20] for finding the most violated cycle inequality in the primal, but is not based on shortest paths. We use the cycle inequalities only as a means of obtaining an efficient search algorithm, and we add the full cycle cluster to the dual decomposition as in previous work.

Somewhat surprisingly, in Section 4 we show that, for *binary* graphical models and when the dual has been solved to optimality, the two bound criterions (the one given in [21] and this one) coincide. Thus, at least for binary graphical models, the algorithm that we present completely solves the open problem from [21] of how to efficiently search over cycle clusters.

## 2 Background

We consider MAP inference problems on factor graphs with $n$ variables $X_1, \ldots, X_n$, where each variable takes discrete states $x_i \in \text{Vals}(X_i)$. The MAP inference problem is then

$$\text{MAP}(\theta) = \max_{\mathbf{x}} \sum_{\mathbf{c} \in C} \theta_{\mathbf{c}}(\mathbf{x_c}), \quad (1)$$

where $C$ denotes the set of factors, and $\theta_{\mathbf{c}}(\mathbf{x_c})$ is the log of the potential function for factor $\mathbf{c}$. Let $G = (V, E)$ be the Markov network corresponding to this factor graph, with one edge $ij \in E$ for every two variables $i$ and $j$ that appear together in some factor. The notation $N(i)$ refers to the set of variables that neighbor $i$ in the Markov network.

### 2.1 Dual Decomposition

The dual decomposition approach [19] attempts to address the intractability of the joint maximization in Eq. 1 by introducing dual variables $\delta$ and minimizing an upper bound on the MAP assignment:

$$\min_{\delta} L(\delta), \quad L(\delta) = \sum_{\mathbf{c} \in C} \max_{\mathbf{x_c}} \theta^{\delta}_{\mathbf{c}}(\mathbf{x_c}), \quad (2)$$

where the reparameterizations $\theta^{\delta}_{\mathbf{c}}(\mathbf{x_c})$ are given by

$$\begin{aligned}
\theta^{\delta}_i(x_i) &= \theta_i(x_i) + \sum_{j \in N(i)} \delta_{ij \to i}(x_i) & \forall i \in V, \\
\theta^{\delta}_{ij}(x_i, x_j) &= \theta_{ij}(x_i, x_j) - \delta_{ij \to i}(x_i) - \delta_{ij \to j}(x_j) \\
&\quad + \sum_{\mathbf{c}: |\mathbf{c}| \geq 3, \, i,j \in \mathbf{c}} \delta_{\mathbf{c} \to ij}(x_i, x_j) & \forall ij \in E, \\
\theta^{\delta}_{\mathbf{c}}(\mathbf{x_c}) &= \theta_{\mathbf{c}}(\mathbf{x_c}) - \sum_{i,j \in \mathbf{c}} \delta_{\mathbf{c} \to ij}(x_i, x_j) & \forall |\mathbf{c}| \geq 3.
\end{aligned}$$

The key property of the function $L(\delta)$ is that it only involves maximization over local assignments $\mathbf{x_c}$, a task which we assume to be tractable. The dual thus decouples the original problem, resulting in a problem that can be optimized using local operations.

If we ever find a global assignment $\mathbf{x}$ which locally maximizes all of the subproblems, then $\mathbf{x}$ is guaranteed to be the MAP assignment. Thus, the dual solution $\delta$ has the ability to provide a *certificate of optimality*. Even in cases when the relaxation is loose, the dual provides an upper bound, $\text{MAP}(\theta) \leq L(\delta)$, that can be useful within branch-and-bound.

The algorithms described in this paper make use of the reparameterized edge potentials $\theta^{\delta}_{ij}(x_i, x_j)$ when searching for frustrated cycles. For notational clarity and to be consistent with earlier work, we use

$b_{ij}(x_i, x_j)$ to denote $\theta_{ij}^\delta(x_i, x_j)$ for the current dual variables $\delta$, also calling these the edge "beliefs".

## 2.2 Cluster Pursuit

If there still remains an integrality gap after solving the current dual, i.e. $\text{MAP}(\theta) < L(\delta^*)$, one can *tighten* the relaxation by introducing new zero-valued potentials $\theta_\mathbf{c}(\mathbf{x_c})$ for clusters $\mathbf{c}$ not originally in the factor graph [21, 24]. The advantage of doing this together with dual decomposition is that one can *warm start*, initializing the new dual's variables using the previous dual solution. The resulting algorithm alternates between minimizing the current dual for some number of iterations (not necessarily to optimality) and searching for new clusters to use in tightening the relaxation. If the dual is solved using coordinate-descent, then every step of the algorithm monotonically improves the upper bound on the MAP value.

The key problem addressed by earlier work was how to choose *which* clusters to use to tighten the relaxation. In particular, Sontag et al. [21] proposed to evaluate the utility of adding a cycle $C$ to the relaxation by the greedy bound minimization criterion

$$d(C) = \sum_{e \in C} \max_{\mathbf{x}_e} b_e(\mathbf{x}_e) - \max_{\mathbf{x}_C} \left[ \sum_{e \in C} b_e(\mathbf{x}_e) \right]. \quad (3)$$

Note that only the edges in the cycle $C$ are used in the maximization over $\mathbf{x}_C$. As a result, $d(C)$ can be computed in $O(k^3|C|)$ time, where $k$ is the number of states for each variable. The criterion corresponds to the amount that the dual would decrease if we add this cycle cluster and perform one block coordinate descent step on all dual variables corresponding to the new cluster. Once a cycle is selected for addition, [21] triangulates the cycle and adds each of the triplet clusters; we do this too.

## 2.3 Linear Programming Relaxation

The dual decomposition for MAP inference given in Eq. 2 can be shown to be equivalent, by LP duality, to the following linear programming relaxation:

$$\max_{\boldsymbol{\mu} \in M_L} \sum_\mathbf{c} \sum_{\mathbf{x_c}} \theta_\mathbf{c}(\mathbf{x_c}) \mu_\mathbf{c}(\mathbf{x_c}) \quad (4)$$

where the *local marginal polytope* $M_L$ enforces that $\{\mu_i(x_i), \forall x_i\}$ and $\{\mu_\mathbf{c}(\mathbf{x_c}), \forall \mathbf{x_c}\}$ correspond to valid (local) probability distributions and that, for each factor $\mathbf{c}$, $\mu_\mathbf{c}(\mathbf{x_c})$ is consistent with $\mu_i(x_i)$ for all $i \in \mathbf{c}, x_i$:

$$M_L = \left\{ \boldsymbol{\mu} \geq 0 : \begin{array}{l} \sum_{x_i} \mu_i(x_i) = 1, \quad \forall i \\ \sum_{\mathbf{x_{c \setminus i}}} \mu_\mathbf{c}(\mathbf{x_c}) = \mu_i(x_i) \\ \sum_{\mathbf{x_{c \setminus \{i,j\}}}} \mu_\mathbf{c}(\mathbf{x_c}) = \mu_{ij}(x_i, x_j) \end{array} \right\},$$

where the second set of constraints is for all $\mathbf{c}, i \in \mathbf{c}, x_i$, and the third set of constraints is for all factors $\mathbf{c}$ such that $|\mathbf{c}| \geq 3$, $i, j \in \mathbf{c}, x_i, x_j$.

The exact MAP inference problem would be obtained if we instead had maximized Eq. 4 over the *marginal polytope* [22], which enforces that all feasible points $\boldsymbol{\mu}$ correspond to marginals arising from some exponential family distribution with the same sufficient statistics. For most graphical models the marginal polytope is intractable to optimize over, which is why we use the relaxation $M_L$.

## 2.4 $k$-ary Cycle Inequalities

The main contribution of this paper is to show how to solve the search problem for cycle consistency constraints (within the framework of dual decomposition) by introducing a new bound criterion based on the $k$-ary cycle inequalities [20].

The cycle inequalities [1, 2, 4] are a set of constraints for the marginal polytope of graphical models with binary variables, which arises from the observation that a cycle must have an even (possibly zero) number of cut edges. Suppose we start at node $i$, where $x_i = 0$. As we traverse the cycle, the assignment changes each time we cross a cut edge. Since we must return to $x_i = 0$, the assignment can only change an even number of times. This concept can be generalized to obtain the $k$-ary cycle inequalities, which are valid constraints for graphical models with non-binary variables [20].

For each variable $i$, let $\pi_i^q$ be a *partition* of all of its states into two disjoint non-empty sets. Let $\pi_i$ denote the set of all partitions of variable $i$. The $k$-ary cycle inequalities are defined using the *projection graph* $G_\pi = (V_\pi, E_\pi)$, which has one node for each partition in each set $\pi_i$ and all such nodes are fully connected across adjacent variables. That is, we have $V_\pi = \bigcup_{i \in V} \pi_i$, and

$$E_\pi = \{(\pi_i^q, \pi_j^r) \mid (i,j) \in E, q \leq |\pi_i|, r \leq |\pi_j|\}. \quad (5)$$

We obtain a different projection graph depending on the quantity and type of partitions that we choose for each variable. The algorithms in this paper are applicable for *any* projection graph. In our experimental results, we primarily use the $k$-projection graph, which simply has a partition $\pi_i^{x_i} = \{x_i\}$ (versus all other states) for every variable $i$ and state $x_i$. Thus, if every variable takes $k$ states, the projection graph will have $k|V|$ nodes and $k^2|E|$ edges. In the supplementary material we describe an algorithm which finds the optimal partition for each variable with respect to each edge. If using the $k$-projection graph does not succeed in finding a frustrated cycle, we re-run the cycle search method using this expanded set of partitions.

There is one $k$-ary cycle inequality for every cycle $C$ in the projection graph $G_\pi$ and for every set of edges $F \subseteq C$ such that $|F|$ is odd:

$$\sum_{mn \in C \setminus F} \sum_{\substack{x_i, x_j: \\ \pi_i^q(x_i) \neq \\ \pi_j^r(x_j)}} \mu_{ij}(x_i, x_j) + \sum_{mn \in F} \sum_{\substack{x_i, x_j: \\ \pi_i^q(x_i) = \\ \pi_j^r(x_j)}} \mu_{ij}(x_i, x_j) \geq 1$$

where $mn = (\pi_i^q, \pi_j^r) \in E_\pi$. Although there are exponentially many $k$-ary cycle inequalities, [20] showed how the most violated inequality can be found in polynomial time, using a shortest-path algorithm. However, unlike [20] we use these inequalities in the *dual*. In the next section, we will give a new algorithm to efficiently find violated $k$-ary cycle inequalities while working within the framework of dual decomposition.

## 3 Cycle Inequalities in the Dual

In this section we give a column-generation approach for adding cycle inequalities within the dual decomposition framework. Consider the terms of the dual objective (Eq. 2) that would be affected by adding a single dual variable $\lambda_{C,F,\pi}$ corresponding to one $k$-ary cycle inequality specified by $C \subseteq E$, $F \subseteq C$ (recall that $|F|$ must be odd), and $\pi$.[1] After adding $\lambda_{C,F,\pi}$, the new dual has the following terms (see supplementary material for derivation):

$$\sum_{ij \in F} \max_{x_i, x_j} \Big( b_{ij}(x_i, x_j) + \lambda_{C,F,\pi} \mathbf{1}[\pi_i(x_i) = \pi_j(x_j)] \Big)$$

$$+ \sum_{ij \in C \setminus F} \max_{x_i, x_j} \Big( b_{ij}(x_i, x_j) + \lambda_{C,F,\pi} \mathbf{1}[\pi_i(x_i) \neq \pi_j(x_j)] \Big)$$

$$- \lambda_{C,F,\pi}.$$

For each edge $ij \in C$, define the weight

$$s_{ij}^\pi = \max_{x_i, x_j : \pi_i(x_i) = \pi_j(x_j)} b_{ij}(x_i, x_j) - \max_{x_i, x_j : \pi_i(x_i) \neq \pi_j(x_j)} b_{ij}(x_i, x_j).$$

---
[1] When used in the context of a cycle in the graphical model, $C \subseteq E$, as opposed to a cycle in the projection graph, the notation $\pi$ specifies a particular partition for each variable along the cycle.

We show in the supplementary material that the coordinate descent step for $\lambda_{C,F,\pi}$ is given by $\lambda_{C,F,\pi} = \min_{ij \in C} w_{ij}^{\pi,F}$, where $w_{ij}^{\pi,F} = s_{ij}^\pi$ for $ij \notin F$ and $w_{ij}^{\pi,F} = -s_{ij}^\pi$ for $ij \in F$.

The amount that the dual objective decreases with one coordinate descent step on $\lambda_{C,F,\pi}$, assuming that $\lambda_{C,F,\pi}$ was previously zero, is

$$d(C, F, \pi) = \max(0, \min_{ij \in C} w_{ij}^{\pi,F}). \quad (6)$$

**Example 1.** Consider a triangle on three edges ($C = \{12, 23, 31\}$), with $x_i \in \{0, 1\}$, where $\theta_i(x_i) = 0$ $\forall i, x_i$ and $\theta_{ij}(x_i, x_j) = 1$ if $x_i \neq x_j$, and 0 otherwise. Since this example is binary, we simply use $\pi_i(x_i) = x_i$. Let all of the dual variables $\delta$ be 0, and assume that initially there are no cycle inequalities. The best integer solution has value 2, while the pairwise LP relaxation gives only a loose upper bound of 3 (note: $\delta$ as defined can be shown to be optimal for the dual, i.e. Eq. 2).

Consider the problem of finding the best cycle inequality according to $\arg\max_{C,F} d(C, F)$. First, note that $b_{ij}(x_i, x_j) = \theta_{ij}(x_i, x_j)$, so $w_{ij}^F = 1$ for $ij \in F$ and $w_{ij}^F = -1$ for $ij \notin F$. If $F = \emptyset$, then $w_{ij}^F = -1$ for all edges, and so $d(C, F) = 0$. On the other hand, if $F = C$, then $w_{ij}^F = 1$ for all edges, which gives a bound decrease of $d(C, F) = 1$, corresponding to $\lambda_{C,F} = 1$.

### 3.1 Separation Algorithm

The bound criterion $d(C, F, \pi)$ given in Eq. 6 is analogous to the bound criterion $d(C)$ used by [21] (see Eq. 3) to evaluate the utility of adding a cycle $C$ to the relaxation. We now consider the algorithmic problem of finding the *best* dual variable $\lambda_{C,F,\pi}$ to add, according to $d(C, F, \pi)$:

$$\max_{C, F \subseteq C \text{ s.t. } |F| \text{ odd}, \pi} d(C, F, \pi). \quad (7)$$

We show that this can be computed efficiently using a variation on the shortest-path based separation algorithm for $k$-ary cycle inequalities [20]. We first note that Eq. 7 is equal to

$$\max \Big( 0, \max_{C, F \subseteq C \text{ s.t. } |F| \text{ odd}, \pi} \min_{ij \in C} w_{ij}^{\pi,F} \Big). \quad (8)$$

We can simplify this further by noticing that for $d(C, F, \pi)$ to be positive, we need $ij \in F$ when $s_{ij}^\pi < 0$, and $ij \notin F$ when $s_{ij}^\pi > 0$. Thus, ignoring the maximization over the partitioning $\pi$, Eq. 8

> **Algorithm FindOddCycle** (Graph $G_\pi$, edge weights $\{s_{mn} : (m,n) \in E_\pi\}$)
> 1. Construct a spanning tree $T$ of $G_\pi$. Set $r$ to be the root of $T$.
> 2. $\text{sign}[r] \leftarrow +1$
> 3. **for** each vertex $t \in T$ (in order of increasing distance from $r$ in $T$):
> 4.      $\text{sign}[t] \leftarrow \text{sign}[\text{pa}(t)] \cdot s_{t,\text{pa}(t)}$
> 5. **for** each $(m,n) \in E_\pi \setminus T$:
> 6.      **if** $\text{sign}[m] \neq \text{sign}[n] \cdot s_{mn}$
> 7.          **return** cycle given by the edge $(m,n)$ and the path $m \rightsquigarrow r \rightsquigarrow n$ in $T$
> 8.
> 9. **return** no odd signed cycle found

Figure 1: Assuming that the edge weights $s_{mn}$ are in $\{-1, 1\}$, this algorithm will find a cycle with an odd number of $-1$ edges, if one exists. $\text{pa}(t)$ refers to the parent of node $t$ in the spanning tree $T$.

is equivalent to

$$\max\left(0, \max_{\substack{C \subseteq E \text{ s.t.} \\ \prod_{ij \in C} \text{sign}(s^\pi_{ij}) = -1}} \min_{ij \in C} |s^\pi_{ij}|\right). \quad (9)$$

We conclude that the maximum bound decrease is achieved by the cycle in $G$ with an odd number of negative weight edges that maximizes the minimum of the absolute value of the weights along the cycle.

For Markov networks with non-binary variables, we also wish to maximize over the partition for each variable. Let $G_\pi$ denote the projection graph, where the variables and edges are as defined in Eq. 5. All subsequent quantities will use the partitions for the variables specified by the edges $mn = (\pi^q_i, \pi^r_j) \in E_\pi$ in the projection graph. Assign weights to the edges[2]

$$s_{mn} = \max_{x_i, x_j : \pi^q_i(x_i) = \pi^r_j(x_j)} b_{ij}(x_i, x_j) - \quad (10)$$
$$\max_{x_i, x_j : \pi^q_i(x_i) \neq \pi^r_j(x_j)} b_{ij}(x_i, x_j),$$

and remove all edges with $s_{mn} = 0$. The algorithm that we describe next will find the maximum of

$$\max\left(0, \max_{\substack{C \subseteq E_\pi \text{ s.t.} \\ \prod_{mn \in C} \text{sign}(s_{mn}) = -1}} \min_{mn \in C} |s_{mn}|\right). \quad (11)$$

Suppose that $s_{mn} \in \{+1, -1\}$. Then, $\min_{mn \in C} |s_{mn}| = 1$ and the optimization problem simplifies to finding a cycle with an odd number of $-1$ edges. This can be solved in linear time by the algorithm given in Figure 1 (if the graph is not connected, do this for each component).[3]

---
[2]For the $k$-projection graph, this step can be implemented efficiently, taking time $O(k^2|E|)$ to compute all edge weights instead of $O(k^2|E_\pi|)$.

[3]In the case when there is more than one cycle with an odd number of $-1$ edges, the particular cycle that we find depends on the choice of spanning tree. However, the algorithm is always guaranteed to find *some* cycle with an odd number of $-1$ edges, when one exists, regardless of the choice of spanning tree.

Now consider the case of general $s_{mn}$. We can solve the optimization in Eq. 11 by doing a binary search on $|s_{mn}|$. There are only $|E_\pi|$ possible edge weights, so to do this search we first sort the values $\{|s_{mn}| : mn \in E_\pi\}$. At each step, we consider the subgraph $G'$ consisting of all $mn \in E_\pi$ such that $|s_{mn}| > R$, where $R$ is the threshold used in the binary search. We then let $s_{m'n'} = \text{sign}(s_{mn})$ for $m'n' \in G'$, and search for a cycle with an odd number of negative edges using the algorithm described in Figure 1. The binary search will find the largest $R$ such that there is a negative-signed cycle $C \in G'$, if one exists. The best choice of $\lambda_{C,F,\pi}$ is then given by this cycle $C$, the edges $F$ corresponding to the negative-weight edges in $C$, and $\pi$ given by the partitions used in $C$. The total running time is only $O(|E_\pi| \log |E_\pi|)$.

If the cycle that is returned uses each variable only once (i.e., does not consider more than one partition for a single variable), the corresponding cycle and choice of partitions will be optimal for Eq. 8. Otherwise, one can show that the guaranteed bound decrease after one coordinate descent step on the new cycle inequality's dual variable is a constant fraction of that guaranteed by Eq. 8.

When we use this algorithm in the experiments of Section 6, we ignore $F$ and $\pi$, instead fully enforcing cycle consistency for the cycle $C$ that is best according to this bound criterion.

## 4 Theoretical Results

Consider the restricted set of clusters $\mathcal{C}_\text{cycles}(G)$ corresponding to cycles of arbitrary length,

$$\mathcal{C}_\text{cycles}(G) = \left\{C \subseteq E \mid C \text{ forms a cycle in } G\right\}. \quad (12)$$

A natural question is whether it is possible to find the best cycle cluster to add to the relaxation according to the greedy bound minimization criteria $d(C)$

[21]. It is easily shown that $\max_{F,\pi} d(C,F,\pi) \leq d(C)$ for all cycles $C$. Thus, if it were not for computational concerns, searching using $d(C)$ would be optimal. Unfortunately, we show a number of negative results proving that searching for the best cycle according to $d(C)$ is computationally intractable.

We show the following, where $k$ refers to the number of states per node.

1. For $k = 2$, when the beliefs $b_e(\mathbf{x}_e)$ are dual optimal, maximizing Eq. 3 is *equivalent* to finding the best cycle inequality in the dual.[4]

2. For $k = 2$, maximizing Eq. 3 is NP-hard when the beliefs $b_e(\mathbf{x}_e)$ are not dual optimal.

3. For $k > 2$, maximizing Eq. 3 is always NP-hard.

By dual optimal, we mean that the beliefs correspond to a dual optimal solution of the current LP relaxation. Note that, before solving the dual LP to optimality, $b_e(\mathbf{x}_e)$ can be almost anything. For example, we can set $\theta_e(\mathbf{x}_e)$ arbitrarily and consider the separation problem at the first iteration.

**Theorem 4.1.** *When $k = 2$ and the beliefs $b_{ij}(x_i, x_j)$ correspond to a dual optimal solution, $\max_{C \in \mathcal{C}_{cycles}(G)} d(C) = \max_{C, F: |F| \text{ odd}} d(C, F)$.*

The proof of Theorem 4.1 makes use of the assumption that $b_{ij}(x_i, x_j)$ corresponds to a dual optimal solution in only one step, when applying the complementary slackness conditions for the edge variables. Thus, Theorem 4.1 holds for any dual decomposition for MAP which is at least as tight as the pairwise LP relaxation. In particular, adding cycle constraints does not change the premise of the theorem.

One conclusion that is immediate given Theorem 4.1 is that (for binary MRFs) the cycle inequalities give at least as tight of a relaxation as the cycle relaxation. In fact, for a single cycle, just one cycle inequality suffices to make the LP relaxation tight for a given instance. This is precisely what we observed in Example 1.

### 4.1 NP-Hardness Results

It is often possible to obtain much better results by tightening the relaxation even before the dual is solved to optimality [21]. Unfortunately, although we showed in Section 3 that for $k$-ary cycle inequalities $\max_{C,F,\pi} d(C,F,\pi)$ can be computed efficiently, the corresponding problem for cycle consistency constraints is significantly more difficult:

---

[4]This result is for binary variables only, so we let the projection be $\pi_i(x_i) = x_i$ and omit the $\pi$ notation.

**Theorem 4.2.** *The optimization problem $\max_{C \in \mathcal{C}_{cycles}(G)} d(C)$ is NP-hard for $k = 2$ and beliefs $b_{ij}(x_i, x_j)$ arising from a non-optimal dual feasible point.*

We next show that, for $k > 2$, not even dual optimality helps:

**Theorem 4.3.** *The optimization problem $\max_{C \in \mathcal{C}_{cycles}(G)} d(C)$ is NP-hard for $k \geq 3$ even for beliefs $b_{ij}(x_i, x_j)$ corresponding to a dual optimal solution of the pairwise relaxation.*

Both proofs use a reduction from the Hamiltonian cycle problem. Full proofs can be found in the supplementary material.

## 5 Related Work

Our new algorithm is closely related to two earlier approaches for tightening the dual using cycle constraints. First, Komodakis et al. proposed to tighten the pairwise LP relaxation by a sequence of cycle repairing operations [12]. Their algorithm is applicable to graphical models with non-binary variables. For binary graphical models, when the dual is at optimality, two cycle repairs – corresponding to the two anchors of any variable (using their terminology) – can be seen to be equivalent to one coordinate descent step on a new cycle inequality for this cycle. We solve the open problem of how to *find* the cycles where cycle repairs are necessary. In their experiments, [12] explicitly enumerated over short cycles.

Second, Johnson proposed an algorithm to find *inconsistent* cycles in the dual [9]. His approach applies only to binary-valued graphical models, and only when the dual is close to optimality. In these cases, his algorithm can be shown to find a cycle $C$ such that $d(C, F) > 0$ for some $F \subseteq C, |F|$ odd. His algorithm, which inspired our approach, constructs $s_{ij} \in \{+1, -1\}$ and looks for inconsistent cycles using the linear time method described in Section 3.1. Because of numerical difficulties, Johnson needed to use an edge-wise correlation measure, computed using the primal solution obtained from the smoothed dual [9, p.134]. By drawing the connection to cycle inequalities, we obtain a *weighted* approach whereas his was *unweighted*. As a result, there are no numerical difficulties, and our algorithm can be applied long before solving the dual to optimality.

It may seem surprising that the dual separation algorithm is so much faster than the primal separation algorithm for $k$-ary cycle inequalities [20]. However, this is because the dual searches over a smaller class of cycle inequalities. Consider the case where we have solved the dual to optimality for the current relaxation. Then, using the complementary slackness conditions one can show that, for any cycle inequality $C, F, \pi$ such that $d(C, F, \pi) > 0$, and for any primal solution $\mu$,

$$\sum_{mn \in C \setminus F} \left( \mu_{mn}^\pi(0,1) + \mu_{mn}^\pi(1,0) \right) \qquad (13)$$
$$+ \sum_{mn \in F} \left( \mu_{mn}^\pi(0,0) + \mu_{mn}^\pi(1,1) \right) = 0.$$

The right hand side could be anything less than 1 for us to obtain a violated cycle inequality, and it is always non-negative because $\mu_{ij}(x_i, x_j) \geq 0$. But, since the right hand side of Eq. 13 is 0, we conclude that the dual separation algorithm is only able to find cycle inequalities to add to the relaxation that are *very violated*. By Theorem 4.1, the same conclusion holds for binary graphical models for the cluster-pursuit algorithm given in [21], when applied to a dual optimal solution – if a triplet cluster $C$ is found such that $d(C) > 0$, then there exists a cycle inequality $C, F$ that is very violated by all primal solutions. It is also possible to give a $O(|E_\pi|)$ time separation algorithm in the primal that would separate this smaller class of $k$-ary cycle inequalities.

## 6 Experiments

In this section we report experimental results for the new cycle search algorithm (we call this the "cycle" method) applied to MAP inference problems arising from predicting protein-protein interactions [5, 8], protein side-chain placement [25], and stereo vision [25]. We compare the cycle method to cluster-pursuit using dictionary enumeration with triplet clusters [21], which we call the "triplet" method. Both algorithms were implemented using C++, and differ only in the mechanism used for cycle search. All experiments were performed on a 2.4 GHz AMD Opteron(tm) machine with 128 GB of memory.

We use the Max-Product Linear Programming (MPLP) algorithm to minimize the dual [6]. On all problems, we start by running MPLP for 1000 iterations. If the integrality gap is less than $10^{-4}$, the algorithm terminates. Otherwise, we alternate between further tightening the relaxation (using either the cycle or triplet method) and running another 20 MPLP iterations, until the problem is solved optimally or a time limit is reached. For the triplet algorithm, we add between 5 and 20 triplets per outer iteration, and for the cycle algorithm, we add from 1 to 5 cycles per outer iteration (we required at least 5 new triplet clusters).

In our experiments we also consider a combination of the two approaches, which we call the "triplet+cycle" method, where we add the clusters found by *both* cycle search algorithms.

**Biasing toward short cycles.** Whereas previous cluster-pursuit approaches considered clusters of only three or four variables, our algorithm could potentially choose to add a cycle involving a large number of variables. Consequently, the *length* of the cycle added can significantly affect the per-iteration running time. Our algorithm must balance between (a) choosing cycles which most improve the dual objective and (b) keeping the per-iteration running time reasonable.

The length of the cycles found depends both on the initial *choice of the spanning tree* (line 1 of algorithm FindOddCycle, given in Figure 1) and on *which edge is chosen* to create the cycle (lines 5–7). Notice that the length of any cycle returned by FindOddCycle is at most twice the depth of the spanning tree $T$. We found that using breadth-first search to create the spanning tree results in much more shallow trees, and as a result significantly shorter cycles compared to depth-first search.

After creating the spanning tree and propagating the signs, any edge that has opposite signs on its endpoints could be chosen to create the cycle (lines 5–7). The cycle corresponds to the union of the edge and the paths from each endpoint to their least common ancestor (LCA) in $T$. We find the lengths of each of these cycles by running a LCA algorithm (we note that this can be done in amortized constant time using the algorithm of [17]). Then, we sort the edges in increasing order according to the length of the corresponding cycles, and return the shortest few.

We also experimented with an algorithm which is guaranteed to find the shortest cycles, but found that its running time was too long to be practical.

### 6.1 Protein-Protein Interaction

We consider 8 inference problems arising from the relational classification task of protein-protein interaction prediction [5, 8]. This inference problem aims at predicting protein-protein interactions (PPIs) given high-throughput protein-protein interaction and other cellular experimental data. The PPI problems are Markov networks with over 14,000 binary variables denoting the cellular localization of the proteins and whether any particular pair of pro-teins interact. The Markov network has over 30,000 node and triad potentials. These inference problems are difficult to solve because the triad potentials induce many frustrated cycles. Each of the 8 inference problems corresponds to the parameters found at some iteration during learning, and the difficulty of inference varies substantially among them.

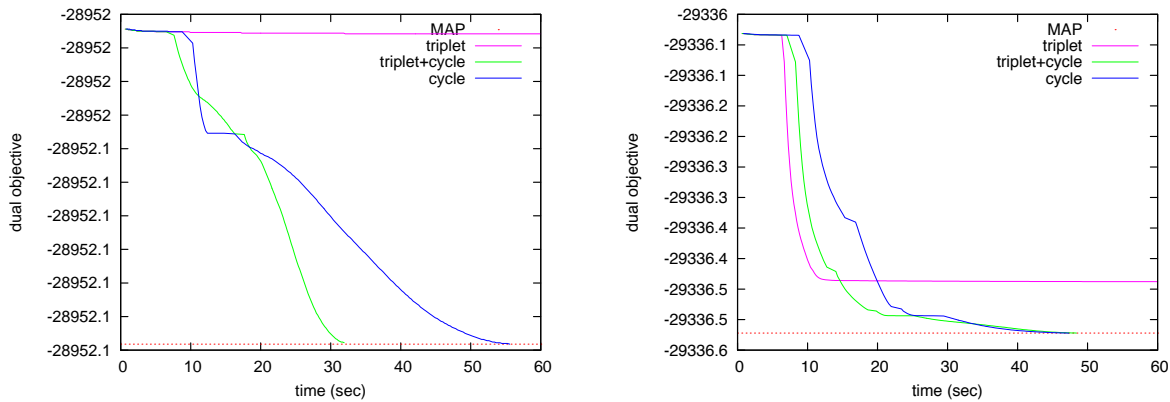

Figure 2: Comparison of the cycle search algorithms on a MAP inference problem from protein-protein interaction prediction. The time shown is for the tightening stage, *after* the initial 1000 iterations of ADLP.

For 2 of the PPI problems, the cycle method finds the MAP assignment in less than a minute after solving the initial dual to optimality (see Figure 2).[5] For the other 6 problems, the cycle method obtains a slightly better dual objective value than the triplet method when we terminated at half an hour.

In contrast, the triplet method is unable to exactly solve any of the PPI problems. We noticed that the bound criterion [21] for most triplets found in these problems is close to 0, which explains the considerable difficulty for the triplet algorithm to quickly choose the best clusters to add to the relaxation.

These results indicate that the cycle algorithm has a significant advantage over the triplet algorithm on inference problems on sparse graphs.

### 6.2 Side-Chain Prediction

We next considered 30 protein side-chain placement inference problems, corresponding to the 30 proteins from [25] for which the pairwise LP relaxation is not tight. Previous work has shown that the triplet method exactly solves these [21]. The triplet algorithm's running time ranges from 1.12 to 182.18 seconds with a median of 20.42 seconds.

---
[5]For the PPI problems MPLP converges very slowly, so we instead use the ADLP algorithm [16] to solve the initial dual, before tightening.

The cycle method also finds the MAP assignment for all 30 proteins. The cycle algorithm's running time is between 2.92 and 3788 seconds, with a median of 24.83 seconds. The outlier which took 3788 seconds to solve was '1kmo'. The cycle method performs significantly worse on this example than the triplet method because the former needs 94 tightening iterations to find the single cluster that solves MAP exactly, whereas the latter finds it immediately.

The cycle algorithm using *only* the $k$-projection graph does not exactly solve '1qb7' and '1rl6'.[6] We show in Figure 3 the dual objective for '1qb7'. At around 50 seconds the cycle method, not finding any useful cycle with the $k$-projection graph, runs the partition finding algorithm explained in the supplementary section, finds the necessary cluster, and solves the protein to optimality.

The "triplet+cycle" method solves all of the side-chain placement problems quickly. As seen in Figure 3, it is slightly slower than the triplet method because of the additional time for running the cycle search method. From these results, we conclude that it is best to use both search methods, when feasible.

### 6.3 Importance of bound criterion

As we mentioned in the related work, for graphical models with binary variables, our cycle search method is similar to the algorithm earlier proposed by Johnson [9]. The key difference is that our approach has a bound criterion which is used to select *which* frustrated cycle to include, whereas Johnson's approach was unweighted (finding *any* odd-signed cycle). In this section we illustrate the importance of having the weighted bound criterion. In addition to the 30 protein side-chain placement problems that we solved exactly, we also consider 4 inference problems arising from stereo vision (all variants of the "Tsukuba" image sequence), which were previously studied by [21, 25].

---
[6][20] also needed the full projection graph for '1rl6'.

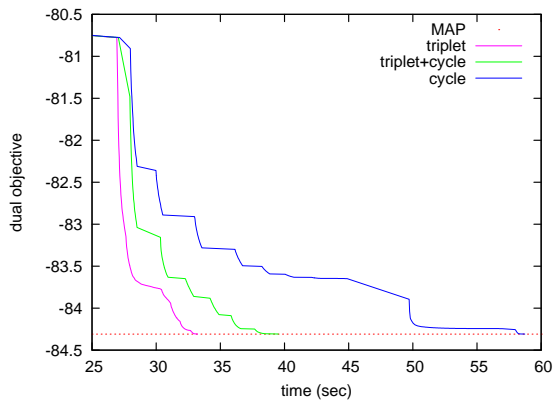

Figure 3: Comparison of all 3 methods on the protein-side chain placement problem '1qb7'.

After finding the optimal value of the bound criterion, $R$ (the threshold used in the binary search, described in Section 3.1), we prune the projection graph to consider only those edges with $|s_{mn}| \geq R/c$ for $c = 1, 4$, and $128$. $c = 1$ corresponds to the usual cycle method, whereas $c = 128$ is more similar to an unweighted bound criterion (analogous to [9]).

We considered a protein failed if we could not obtain a certificate within 15 minutes. On the protein side-chain problems, we found that $c = 4$ worked best, solving all proteins but '1kmo' within 3.2 minutes. With $c = 1$, the cycle method additionally failed to solve '1ug6'. With $c = 128$, the cycle method additionally failed to solve '1a8i', '1et9' and '1gsk' (5 failures in total). One reason why $c = 4$ may be preferable over $c = 1$ for non-binary problems is because – since more edges are considered – the cycle method is more likely to find a shorter length cycle which uses each variable only once, rather than multiple times with different partitions.

When testing on stereo vision data, we found that the cycle algorithm was able to solve all 4 problems exactly within 1 hour with $c = 1$, but could not solve any of them within 3 hours for $c = 128$. We conclude that the bound criterion is essential for the cycle method to quickly solve MAP inference problems.

After observing that varying the number of MPLP iterations in each of the outer loops (i.e., after adding cycles) had a very minor effect on the overall running times, we also conclude that the cycle method is robust to the dual not being solved to optimality.

### 6.4 Summary of results

Using the "triplet+cycle" method, we exactly and quickly solved all 30 protein side-chain problems and 2 protein-protein interaction problems.

The cycle algorithm solved all 4 stereo problems exactly within 24–48 minutes using only 9–33 tightening iterations. The number of clusters added by the algorithm were between 227 and 1217.

## 7 Discussion

We believe that our algorithm for finding frustrated cycles may also be useful in graphical models that are not sparse. In these settings, dictionary enumeration methods which tighten the relaxation one triplet cluster at a time can eventually succeed at making all cycles consistent. However, the cycle method has several theoretical advantages. First, the dual bound criterion can be zero for some triplet clusters, but non-zero for a larger cycle involving the same variables; we give a concrete example of this in the supplementary material. In these cases, the cycle method would succeed at improving the dual objective, whereas the triplet method would obtain no guidance from the bound criterion and as a result would randomly choose a triplet to use in tightening the relaxation. Second, if a frustrated cycle is long, the triplet method – which adds clusters one at a time – would take several iterations before eventually adding all triplet clusters which triangulate the cycle, thus enforcing cycle consistency. In contrast, the cycle method would take a single iteration.

In this paper, we used the dual variables corresponding to the cycle inequalities only as a means for finding a frustrated cycle, after which we fully enforce cycle consistency. When the variables have a large number of states, enforcing even one cycle consistency constraint can significantly increase the computation time [18]. In these situations, it may be preferable to instead directly do coordinate descent with respect to the cycle inequality dual variables.

Finally, although in this paper we used the dual algorithm as a stand-alone MAP solver, for some problems it may be more effective to use within branch-and-bound, giving a branch-and-cut approach. The efficient cycle search algorithm can also be used together with any method for solving the dual decomposition, such as ADLP [16].

# References

[1] F. Barahona. On cuts and matchings in planar graphs. *Mathematical Programming*, 60:53–68, 1993.

[2] F. Barahona and A. R. Mahjoub. On the cut polytope. *Mathematical Programming*, 36:157–173, 1986.

[3] D. Batra, S. Nowozin, and P. Kohli. Tighter relaxations for MAP-MRF inference: A local primal-dual gap based separation algorithm. *Journal of Machine Learning Research - Proceedings Track*, 15:146–154, 2011.

[4] M. M. Deza and M. Laurent. *Geometry of Cuts and Metrics*, volume 15 of *Algorithms and Combinatorics*. Springer, 1997.

[5] G. Elidan, I. Mcgraw, and D. Koller. Residual belief propagation: informed scheduling for asynchronous message passing. In *UAI*, 2006.

[6] A. Globerson and T. Jaakkola. Fixing max-product: Convergent message passing algorithms for MAP LP-relaxations. In J. Platt, D. Koller, Y. Singer, and S. Roweis, editors, *NIPS*. MIT Press, Cambridge, MA, 2007.

[7] T. Hazan and A. Shashua. Norm-product belief propagation: primal-dual message-passing for approximate inference. *IEEE Trans. Inf. Theor.*, 56(12):6294–6316, Dec. 2010.

[8] A. Jaimovich, G. Elidan, H. Margalit, and N. Friedman. Towards an integrated protein-protein interaction network: A relational Markov network approach. *Journal of Computational Biology*, 13(2):145–164, 2006.

[9] J. Johnson. *Convex Relaxation Methods for Graphical Models: Lagrangian and Maximum Entropy Approaches*. PhD thesis, EECS, MIT, 2008.

[10] V. Jojic, S. Gould, and D. Koller. Fast and smooth: Accelerated dual decomposition for MAP inference. In *Proceedings of International Conference on Machine Learning (ICML)*, 2010.

[11] V. Kolmogorov. Convergent tree-reweighted message passing for energy minimization. *IEEE Trans. Pattern Anal. Mach. Intell.*, 28(10):1568–1583, 2006.

[12] N. Komodakis and N. Paragios. Beyond loose LP-relaxations: Optimizing MRFs by repairing cycles. In *ECCV*, pages 806–820, 2008.

[13] N. Komodakis, N. Paragios, and G. Tziritas. MRF energy minimization and beyond via dual decomposition. *Pattern Analysis and Machine Intelligence, IEEE Transactions on*, 33(3):531–552, March 2011.

[14] V. A. Kovalevsky and V. K. Koval. A diffusion algorithm for decreasing energy of max-sum labeling problem. Glushkov Institute of Cybernetics, Kiev, USSR. Unpublished, approx. 1975.

[15] M. Laurent. A comparison of the Sherali-Adams, Lovász-Schrijver, and Lasserre relaxations for 0–1 programming. *Math. Oper. Res.*, 28(3):470–496, 2003.

[16] O. Meshi and A. Globerson. An alternating direction method for dual MAP LP relaxation. In *Proceedings of ECML PKDD*, pages 470–483, Berlin, Heidelberg, 2011. Springer-Verlag.

[17] B. Schieber and U. Vishkin. On finding lowest common ancestors: simplification and parallelization. *SIAM J. Comput.*, 17(6):1253–1262, Dec. 1988.

[18] D. Sontag, A. Globerson, and T. Jaakkola. Clusters and coarse partitions in LP relaxations. In *NIPS 21*, pages 1537–1544. MIT Press, 2009.

[19] D. Sontag, A. Globerson, and T. Jaakkola. Introduction to dual decomposition for inference. In S. Sra, S. Nowozin, and S. J. Wright, editors, *Optimization for Machine Learning*. MIT Press, 2011.

[20] D. Sontag and T. Jaakkola. New outer bounds on the marginal polytope. In *NIPS 20*, pages 1393–1400, Cambridge, MA, 2008. MIT Press.

[21] D. Sontag, T. Meltzer, A. Globerson, Y. Weiss, and T. Jaakkola. Tightening LP relaxations for MAP using message-passing. In *UAI*, pages 503–510. AUAI Press, 2008.

[22] M. Wainwright and M. I. Jordan. *Graphical Models, Exponential Families, and Variational Inference*. Now Publishers Inc., Hanover, MA, USA, 2008.

[23] T. Werner. A linear programming approach to max-sum problem: A review. *IEEE Trans. Pattern Anal. Mach. Intell.*, 29(7):1165–1179, 2007.

[24] T. Werner. High-arity interactions, polyhedral relaxations, and cutting plane algorithm for soft constraint optimisation (MAP-MRF). In *CVPR*, 2008.

[25] C. Yanover, T. Meltzer, and Y. Weiss. Linear programming relaxations and belief propagation – an empirical study. *JMLR*, 7:1887–1907, 2006.